\pdfoutput=1

\documentclass[aps,prl,reprint,groupedaddress,footinbib]{revtex4-1}
\usepackage{graphicx} %
\usepackage{amsmath}
\usepackage{mathtools} %
\usepackage{bm}
\usepackage{amssymb} %

\usepackage{hyperref}
\hypersetup{colorlinks=true,linkcolor=blue,citecolor=blue,urlcolor=blue}

\usepackage{braket} %

\newcommand{\eu}{\mathrm{e}^}
\newcommand{\iu}{\ensuremath{\mathrm{i}}}
\newcommand{\rmd}{\mathrm{d}}
\newcommand{\half}{{\ensuremath{\frac{1}{2}}}}

\newcommand{\op}[1]{\ensuremath{\hat{#1}}}
\providecommand{\mat}[1]{\mathsf{#1}}
\renewcommand{\mathbf}[1]{\bm{#1}}
\newcommand{\del}{\mathbf{\nabla}}
\DeclareMathOperator{\Tr}{Tr}

\DeclareMathOperator{\Imag}{Im}

\renewcommand{\Im}{\Imag}

\newcommand{\ImF}{\mbox{$\Imag F$}}

\newcommand{\der}[3][]{\frac{\rmd^{#1}{#2}}{\rmd{#3}^{#1}}}
\newcommand{\pder}[3][]{\frac{\partial^{#1}{#2}}{\partial{#3}^{#1}}}
\newcommand{\pders}[3]{\frac{\partial^2{#1}}{\partial{#2}\partial{#3}}}

\newcommand{\Eqn}[1]{Equation\,(\ref{#1})}
\newcommand{\eqn}[1]{Eq.\,(\ref{#1})}

\newcommand{\fig}[1]{Fig.\,\ref{fig:#1}}

\newcommand{\Ref}[1]{Ref.~[\onlinecite{#1}]}
\newcommand{\Refs}[1]{Refs.~\cite{#1}} %

\begin{document}

\title{Derivation of instanton rate theory from first principles}

\author{Jeremy O. Richardson}
\email{jeremy.richardson@durham.ac.uk}
\affiliation{Department of Chemistry, Durham University,
Durham, DH1 3LE, United Kingdom}

\date{\today}

\begin{abstract}
\noindent
Instanton rate theory is used to study
tunneling events in a wide range of systems
including low-temperature chemical reactions.
Despite many successful applications,
the method
has never been obtained from first principles,
relying instead on the ``\ImF'' premise.
In this paper, the same expression for the
rate of barrier penetration at finite temperature
is rederived %
from quantum scattering theory
[W. H. Miller, S. D. Schwartz, and J. W. Tromp, J. Chem. Phys. \textbf{79}, 4889 (1983)]
using a semiclassical Green's function formalism.
This justifies the instanton approach
and provides a route to deriving the rate of other processes.
\end{abstract}

\pacs{
03.65.Sq %
03.65.Xp %
82.20.Ln, %
82.20.Xr %
} 

\maketitle

\noindent
Nuclear tunneling can significantly affect chemical reactivity \cite{BellBook,Carpenter2011tunnelling,Ley2012tunnelling},
but the most common theoretical methods for
estimating reaction rates \cite{Pechukas1981TST,Truhlar1983TST,*Truhlar1996TST,Haenggi1990rate}
treat the nuclear dynamics using classical principles,
which neglect these important effects.
In large complex systems,
quantum dynamics is far more difficult to simulate than its classical counterpart.
However, using semiclassical considerations,
one can describe certain quantum effects %
with an efficiency similar to that of a classical calculation.
Here, a first-principles derivation is presented for semiclassical instanton theory
which describes
the rate of quantum-mechanical tunneling through an energy barrier,
such as occur in low-temperature chemical reactions.

Despite the wide use of instanton rate theory in various scientific disciplines
from subnuclear physics 
to cosmology \cite{Uses_of_Instantons,Benderskii,Caldeira1983dissipation,Siebrand1999AIM,QuantumGravity},
its derivation is not well understood.
The traditional route is based on the premise
that the rate, $k$, 
is related to the system's free-energy, $F$, by
$k\approx - (2/\hbar) \Im F$
\cite{Langer1967ImF,*Langer1969ImF,Stone1977ImF,Coleman1977ImF,*Callan1977ImF},
and its application to finite-temperature reactions \cite{Benderskii} is understood simply
as an approximate interpolation between known low and high-temperature limits \cite{Affleck1981ImF}.
The imaginary part of an energy is not a well-defined concept,
especially in a bound system \cite{Gillan,Aoyama1997instanton}.
It is obtained by conjecture \cite{Langer1967ImF}
using an analytic continuation of a divergent integral \cite{Kleinert}.
An alternative (and earlier) formulation of instanton theory by Miller 
\cite{Miller1975semiclassical,Chapman1975rates}
employs the heuristic Weyl correspondence rule \cite{Miller1974QTST}
in a
transition-state theory (TST) approximation \cite{Miller1975semiclassical}.
This gives, as an intermediate step in the derivation,
an expression first given by Wigner
\cite{Wigner1932parabolic},
which is not valid \cite{Hele2013QTST} at the low temperatures 
where the instanton is applied.
In both cases, however, semiclassical approximations to the expressions
result in the same instanton rate \cite{Althorpe2011ImF}.

Recently it has become possible to evaluate
these tunneling rates in complex molecular systems
using the ring-polymer instanton (RPI) method \cite{rpinst}.
This approach locates the instanton on the full potential-energy surface
by searching for stationary points of the discretized action
using multidimensional optimization techniques.
It has been applied successfully to many problems of interest
from reactive scattering to diffusion on metal surfaces and hydrogen transfers in enzymes
\cite{Andersson2009Hmethane,*Andersson2011HCO,*Jonsson2011surface,
	DMuH,
	Goumans2010Hbenzene,*Goumans2011Hmethanol,*Meisner2011isotope,*Rommel2011locating,*Rommel2011grids,*Rommel2012enzyme,*Kaestner2013carbenes,*Kaestner2014review}.
Other related approaches are also based on \ImF\
\cite{Makarov1995QTST,Cao1996QTST,Mills1997QTST,
	Kryvohuz2011rate,*Kryvohuz2012abinitio,*Kryvohuz2012instanton,*Kryvohuz2013derivation,*Kryvohuz2014KIE,
	Shushkov2013instanton}.
Note that instanton theory describing tunneling splitting between degenerate minima is not discussed here as its derivation is already rigorous
\cite{Uses_of_Instantons,Benderskii,tunnel,*water,*octamer}.
The RPI method also plays a significant role
in explaining the success of the ring-polymer molecular dynamics (RPMD) method \cite{Habershon2013RPMDreview}
for computing reaction rates in the deep-tunneling regime \cite{rpinst,Hele2013QTST}.
The quantum instanton (QI) approach is also related,
although its applicability
is somewhat hampered by the requirement to locate two optimal dividing surfaces \cite{Miller2003QI,Vanicek2005QI}.

It is well established that the instanton describes the correct physics \cite{Weiss}
and rates compare favorably with exact quantum calculations \cite{Chapman1975rates,Andersson2009Hmethane,DMuH}.
However, despite these successes,
no first-principles derivation of instanton rate theory has been presented up till now.
Here, a formalism is used
based on recently obtained expressions
for semiclassical approximations to the Green's functions
in the classically forbidden region \cite{GoldenGreens}.
The same approach can be used to derive a golden-rule instanton approach
for nonadiabatic electron-transfer reactions \cite{GoldenGreens,GoldenRPI},
and thus unifies the adiabatic (where the Born-Oppenheimer approximation is valid) and nonadiabatic limits of reaction rates into one theory.

Consider the dynamics of an adiabatic chemical reaction.
The Hamiltonian is
$\op{H} = |\op{\mat{p}}|^2/2m + V(\op{\mat{x}})$,
where
$\mat{x}=(x_1,\dots,x_f)$
are the Cartesian coordinates
of $f$ nuclear degrees of freedom.
These nuclei move on the potential-energy surface $V(\mat{x})$
with conjugate momenta $\mat{p}=(p_1,\dots,p_f)$.
Without loss of generality, the degrees of freedom have been mass-weighted such that each has the same mass, $m$.
For simplicity it will be assumed that the Hamiltonian is neither translationally nor rotationally invariant,
but the following arguments can easily be generalized for this case
\cite{Eyring1938rate}.

An $(f-1)$-dimensional dividing surface, defined by $\sigma(\mat{x})=0$,
separates reactants, $\sigma<0$, from products, $\sigma>0$.
The reaction probability at energy $E$ is \cite{Miller1983rate}
\begin{align}
	\label{scattering}
	P(E) &= 2 \hbar^2 \Tr\left[ \op{F} \Im \op{G}(E) \op{F} \Im\op{G}(E) \right],
\end{align}
where $\op{G}(E)=\lim_{\eta\rightarrow0^+}(E+\iu\eta-\op{H})^{-1}$ is the Green's function. 
The flux from reactants to products is \cite{Miller1974QTST,Miller1983rate}
\begin{align}
	\op{F}
	&= \frac{\iu}{\hbar} \left[ \op{H}, \theta[\sigma(\op{\mat{x}})] \right]
	= \frac{\delta[\sigma(\op{\mat{x}})] \, \op{p}_\sigma + \op{p}_\sigma^\dag \, \delta[\sigma(\op{\mat{x}})]}{2m},
\end{align}
where %
$\op{p}_\sigma = \frac{\partial\sigma}{\partial \op{\mat{x}}} \cdot \op{\mat{p}}$
and $\theta$ is the Heaviside step function.
The exact reaction probability is invariant to $\sigma(\mat{x})$ \cite{Miller1983rate}
but it is normally sensible to choose it such that it cuts through the barrier.
It is
\begin{align}
	\label{PE}
	P(E)
	&= \frac{\hbar^2}{m^2} \iint
		\rho(\mat{x}',\mat{x}'')
		\delta[\sigma(\mat{x}')] \delta[\sigma(\mat{x}'')] \rmd\mat{x}' \rmd\mat{x}'',
\end{align}
where %
\begin{multline}
	\rho(\mat{x}',\mat{x}'') = 
		\braket{\mat{x}'|\op{p}_\sigma \Im\op{G}(E)|\mat{x}''}\braket{\mat{x}''|\op{p}_{\sigma} \Im\op{G}(E)|\mat{x}'}
		\\
		+ \braket{\mat{x}'|\op{p}_{\sigma} \Im\op{G}(E) \, \op{p}_{\sigma}^\dag |\mat{x}''}\braket{\mat{x}''|\Im\op{G}(E)|\mat{x}'}.
\end{multline}

The thermal reaction rate, $k$, is given by
\begin{align}
	\label{kthermal}
	k Z_\mathrm{r} = \frac{1}{2\pi\hbar} \int P(E) \, \eu{-\beta E} \, \rmd E,
\end{align}
where $Z_\mathrm{r} = \Tr\left[\eu{-\beta\op{H}}\theta[-\sigma(\op{\mat{x}})]\right]$ is the partition function of the reactants
at reciprocal temperature $\beta=1/k_\mathrm{B}T$.
Assuming %
an appropriate separation of time-scales \cite{ChandlerGreen},
this problem also describes 
the rate of escape from a metastable well
and thus condensed-phase reactions.

The formulation presented so far defines the quantum reaction rate
but cannot be applied to complex systems due to the difficulty of obtaining the exact multidimensional Green's functions.
Instead, they will be treated by
the semiclassical approximation described in \Ref{GoldenGreens},
which gives the asymptotic result in the $\hbar\rightarrow0$ limit \cite{BenderBook}.
This is an extension of Gutzwiller's formulation \cite{Gutzwiller1967semiclassical,*Gutzwiller1971orbits,*GutzwillerBook}
to the classically forbidden region where $V(\mat{x}'),V(\mat{x}'') > E$.
Here the imaginary part of the semiclassical Green's functions
can be written as a sum over imaginary-time classical trajectories
that bounce at a point where $V(\mat{x})=E$.
Imaginary-time trajectories have equations of motion equivalent to Newtonian dynamics in an upside-down potential \cite{Miller1971density}.
Complex-time trajectories that enter the classically allowed region can be ignored,
as these add phase oscillations to the Green's functions
and give a subdominant contribution to the integral in \eqn{kthermal}
 \cite{GoldenGreens}.

\begin{figure}
	\includegraphics{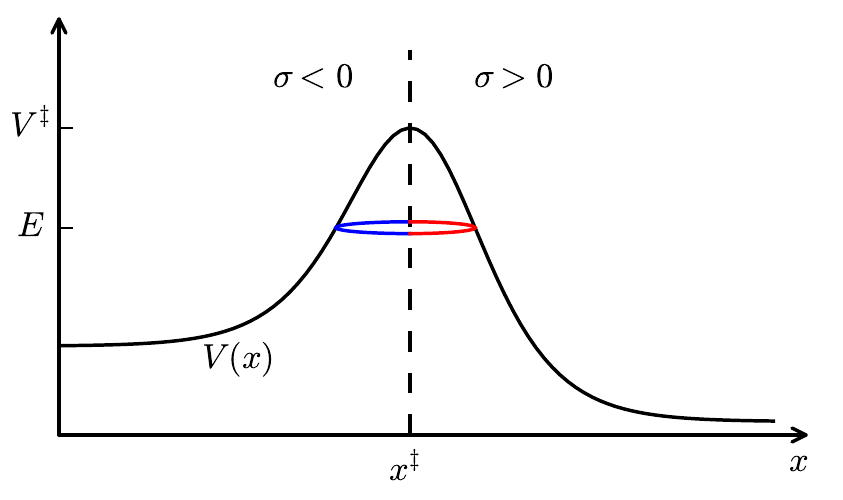}
	\caption{Schematic showing the instanton orbit modeling
	tunneling through a reaction barrier of height $V^\ddag$.
	The orbit is made up of two trajectories %
	that both start and end at the dividing surface $\sigma=x-x^\ddag$ (dashed line)
	but bounce either on the left or right
	and contribute to $\Gamma^-$ or $\Gamma^+$ respectively.
	}
	\label{fig:oneD}
\end{figure}

Only trajectories starting and ending at the dividing surface contribute to \eqn{PE}.
For a tunneling reaction, such as that depicted in \fig{oneD},
where the energy is lower than the barrier height,
there will be two bouncing trajectories that encounter a turning point either on the $+$ or $-$ side of the dividing surface, where $\pm\sigma>0$.
Those that bounce more than once
can be ignored,
as they have larger actions and therefore exponentially smaller contributions.
The imaginary part of the Green's function is then
$\braket{\mat{x}'|\Im\op{G}(E)|\mat{x}''} \simeq
\Gamma^- + \Gamma^+$,
where the contribution from each trajectory is \cite{GoldenGreens}
\begin{align}
	\Gamma^\pm \equiv \Gamma^\pm(\mat{x}',\mat{x}'',E) =
			- \frac{\pi\sqrt{\bar{D}^\pm}}{(2\pi\hbar)^{(f+1)/2}} \, \eu{-\bar{W}^\pm/\hbar}.
\end{align}
The abbreviated action is the following line integral along the respective classical trajectory:
\begin{align}
	\label{Wbar}
	\bar{W}^\pm &\equiv \bar{W}^\pm(\mat{x}',\mat{x}'',E) = \int_{\mat{x}(q)=\mat{x}''}^{\mat{x}(q)=\mat{x}'} \bar{p}(\mat{x}) \, \rmd q, \\
	\bar{p}(\mat{x}) &= \sqrt{2m[V(\mat{x})-E]},
\end{align}
and the prefactors are
\begin{align}
	\bar{D}^\pm &= (-1)^{f+1} \left| \begin{matrix} \pders{\bar{W}^\pm}{\mat{x}'}{\mat{x}''} & \pders{\bar{W}^\pm}{\mat{x}'}{E} \\
										\pders{\bar{W}^\pm}{E}{\mat{x}''} & \pder[2]{\bar{W}^\pm}{E} \end{matrix} \right|
	= \frac{m^2}{\bar{p}(\mat{x}')\bar{p}(\mat{x}'')} A^\pm,
	\label{Dbar}
\\
	A^\pm &= \left| - \pders{\bar{W}^\pm}{\mat{Q}'}{\mat{Q}''} \right|,
\end{align}
where the coordinate system has been transformed from $\mat{x}$ to $(q,\mat{Q})$ \cite{Gutzwiller1971orbits},
defined such that $q$ is parallel to the trajectory and equal to 0 at the dividing surface,
and $\mat{Q}=(Q_1,\dots,Q_{f-1})$ are the perpendicular modes 
\footnote{It will not necessary to use curvilinear coordinates to evaluate the final expressions.
All that will be required 
is that $q$ is parallel at the hopping point $\mat{x}^\ddag$
such that a point transformation suffices.}%
.

The reaction probability, \eqn{PE},
requires not only matrix elements of the Green's function
but also the application of momentum operators on them.
These operators can be written in the position basis as $\op{p}_j=-\iu\hbar\pder{}{\op{x}_j}$,
such that their effect is that of differentiation of the Green's function \cite{Miller1983rate}.
However, because only the terms of the lowest order in $\hbar$ are required for the semiclassical approximation,
the differentiation can be applied only to the exponential.
The operator thus simply multiplies the Green's function by
$\pm\iu\bar{p}(\mat{x}')\pder{x_j'}{q'}$
(or the equivalent with double primes),
which are the momentum components at the end points of the trajectory;
they are imaginary
and the sign depends on the direction traveled.
Within the semiclassical approximation, therefore, the momentum operators act like classical variables.

Using the symmetry of $\Gamma^\pm(\mat{x}',\mat{x}'',E) = \Gamma^\pm(\mat{x}'',\mat{x}',E)$,
\begin{align}
	\rho(\mat{x}',\mat{x}'')
	&\simeq \iu^2 \Big[
	(\bar{p}_\sigma'\Gamma^- - \bar{p}_\sigma'\Gamma^+)
	(\bar{p}_\sigma''\Gamma^- - \bar{p}_\sigma''\Gamma^+)
		\nonumber\\&\qquad
		+ (-\bar{p}_\sigma'\Gamma^-\bar{p}_\sigma'' - \bar{p}_\sigma'\Gamma^+\bar{p}_\sigma'')(\Gamma^- + \Gamma^+) \Big]
\\
	&= 4 \bar{p}_\sigma' \bar{p}_\sigma'' \Gamma^- \Gamma^+,
\end{align}
where $\bar{p}_\sigma'
= \left| \pder{\sigma}{q'} \right| \bar{p}(\mat{x}')$
is the magnitude of the momentum normal to the dividing surface at the end point $\mat{x}'$;
the definition with double primes is equivalent. %
All terms cancel except
the cross term 
with trajectories that bounce once on the left and once on the right.
Unlike for the QI method \cite{Miller2003QI}, it was not necessary to introduce a second dividing surface
to ensure this outcome
\footnote{Although the same expression for $P_\text{SC}(E)$ would be obtained here
using different dividing surfaces for the two fluxes in \eqn{scattering}.
The rate would be independent of each as long
as they are both crossed by the instanton orbit at some point.}.
This is because spurious half-instantons,
which cause Wigner's TST to fail at low temperature \cite{Hele2013QTST},
cannot form as trajectories contributing to $\Im\op{G}(E)$ are required to bounce.

Therefore,
using
$\delta[\sigma(\mat{x})] = \delta(q) \left|\pder{\sigma}{q}\right|^{-1}$\!,
the semiclassical reaction probability is
\begin{multline}
	P_\text{SC}(E)
	= (2\pi\hbar)^{1-f} \!\!
	\iiiint_\text{SD} \! \frac{\bar{p}(\mat{x}')\bar{p}(\mat{x}'')}{m^2} \sqrt{\bar{D}^-\bar{D}^+} \, \eu{-\bar{W}/\hbar}
	\\ \times
	\delta(q') \delta(q'') \, \rmd q' \rmd q'' \rmd\mat{Q}' \rmd\mat{Q}'',
\end{multline}
where
$\bar{W} = \bar{W}^- \! + \bar{W}^+$ is the total action along both trajectories.
Performing the integrals over
$\mat{Q}'$ and $\mat{Q}''$
by the method of steepest descent (SD)
gives
\begin{align}
	P_\text{SC}(E)
	&= Z^\ddag \,
	\eu{-\bar{W}/\hbar}
\\
	Z^\ddag &= 
		\sqrt{A^- A^+}
		\begin{vmatrix} \pders{\bar{W}}{\mat{Q}'}{\mat{Q}'} & \pders{\bar{W}}{\mat{Q}'}{\mat{Q}''} \\
				\pders{\bar{W}}{\mat{Q}''}{\mat{Q}'} & \pders{\bar{W}}{\mat{Q}''}{\mat{Q}''} \end{vmatrix}^{-\half}\!\!.
	\label{Zddag}
\end{align}
All quantities are evaluated at the stationary point $\mat{x}'=\mat{x}''=\mat{x}^\ddag$
on the dividing surface
where $\pder{\bar{W}}{\mat{Q}'}=\pder{\bar{W}}{\mat{Q}''}=\mat{0}$.
Here the trajectories join smoothly into each other to form a continuous periodic orbit,
known as an instanton.

In the one-dimensional case, the formula reduces to $P_\text{SC}(E)=\eu{-\bar{W}/\hbar}$,
which is the well-known WKB result \cite{Bell1935WKB}.
The appendix outlines a proof
that $Z^\ddag$ is a particular generalization of the partition function of the instanton
such that $P_\text{SC}(E)$ is equivalent to an expression given by Miller in 
\Ref{Miller1975semiclassical}.
The final result is therefore independent of the choice of dividing surface
and requires only that the instanton orbit intersects the surface at some point.
The instanton could be thought of as defining a dividing region around the barrier
\cite{Miller1993QTST}.
Note that the short-time approximation inherent in the semiclassical Green's functions
is not necessarily valid when computing microcanonical rates
as it cannot describe nuclear coherences leading, for instance, to discrete densities of states in a reactant well.
The approximation is however asymptotically correct when energy is integrated over a smooth distribution
such as the thermal distribution considered next.

The semiclassical thermal rate is found by evaluating the integral in \eqn{kthermal}
by steepest-descent \cite{Miller1975semiclassical}
to give
\begin{align}
	\label{kSC}
	k_\text{SC} Z_\mathrm{r}
	&= (2\pi\hbar)^{-\half} P_\text{SC}(E) \left(\der[2]{\bar{W}}{E}\right)^{-\half} \eu{-\beta E},
\end{align}
where $E$ solves $\pder{\bar{W}}{E}=\beta\hbar$.
As the imaginary time taken by each trajectory is
$\tau^\pm=-\pder{\bar{W}^\pm}{E}$,
the total time is $\beta\hbar$.
The total derivatives are found using $q'=q''=0$ and recognizing that $\mat{Q}'$ and $\mat{Q}''$ are functions of $E$.

Assuming the barrier
approximates the parabola $V(x)=-m\bar{\omega}^2x^2$ in one degree of freedom near its top,
it cannot support periods less than $2\pi/\bar{\omega}$.
The instanton approach is thus only defined for low temperatures when the periodic orbit exists.
Extensions of the approach to treat higher temperatures,
and involving terms with higher orders of $\hbar$,
have been suggested \cite{Weiss,Kryvohuz2011rate,Zhang2014interpolation}.

The result can be converted to the Lagrangian formulation
using a Legendre transformation
similar to that in \Ref{GoldenGreens}.
This is
based on the full action,
\begin{align}
	\bar{S}^\pm \equiv \bar{S}^\pm(\mat{x}',\mat{x}'',\tau^\pm) = \bar{W}^\pm(\mat{x}',\mat{x}'',E) + E\tau^\pm,
\end{align}
where $E$ is defined such that the trajectories from $\mat{x}''$ to $\mat{x}'$ are completed in imaginary time $\tau^\pm$.
Using $\bar{S}=\bar{S}^-+\bar{S}^+$,
and 
$ \der[2]{\bar{W}}{E} = -\hbar \der{\beta}{E} = -\hbar \big(\der{E}{\beta}\big)^{-1}$\!\!, %
\eqn{kSC} becomes
\begin{align}
	\label{kMiller}
	k_\text{SC} Z_\mathrm{r}
	&= (2\pi\hbar^2)^{-\frac{1}{2}} \, Z^\ddag \left(-\der{E}{\beta}\right)^{\half} \eu{-\bar{S}/\hbar},
\end{align}
which was also obtained by Miller \cite{Miller1975semiclassical},
or equivalently
\begin{align}
	\label{kGreens}
	k_\text{SC} Z_\mathrm{r}
	&= (2\pi\hbar)^{-\half} \sqrt\frac{\Sigma^-\Sigma^+}{-\Sigma} \, \eu{-\bar{S}/\hbar},
\end{align}
where 
$\tau=\tau^+=\beta\hbar-\tau^-$
and, from \Ref{GoldenGreens}, 
\begin{align*}
	&\Sigma^\pm = 
	\begin{vmatrix} \pders{\bar{S}^\pm}{\mat{Q}'}{\mat{Q}''} & \pders{\bar{S}^\pm}{\mat{Q}'}{\tau^\pm} \\
		\pders{\bar{S}^\pm}{\tau^\pm}{\mat{Q}''} & \pders{\bar{S}^\pm}{\tau^\pm}{\tau^\pm} \end{vmatrix}
		= (-1)^{f-1} A^\pm \pder[2]{\bar{S}^\pm}{\tau^\pm},
\\
	&\Sigma =
	\der[2]{\bar{S}}{\tau}
	\begin{vmatrix} \pders{\bar{S}}{\mat{Q}'}{\mat{Q}'} & \pders{\bar{S}}{\mat{Q}'}{\mat{Q}''} \\
		\pders{\bar{S}}{\mat{Q}''}{\mat{Q}'} & \pders{\bar{S}}{\mat{Q}''}{\mat{Q}''} \end{vmatrix}
	= \der[2]{\bar{W}}{E}
		\begin{vmatrix} \pders{\bar{W}}{\mat{Q}'}{\mat{Q}'} & \pders{\bar{W}}{\mat{Q}'}{\mat{Q}''} \\
				\pders{\bar{W}}{\mat{Q}''}{\mat{Q}'} & \pders{\bar{W}}{\mat{Q}''}{\mat{Q}''} \end{vmatrix}.
\end{align*}

\Eqn{kGreens} can be evaluated numerically using
the RPI algorithms to obtain the instanton and its action \cite{rpinst}
and derivatives \cite{GoldenRPI}.
This may lead to a better strategy for evaluating instanton rates
in multidimensional complex systems than the standard RPI approach, \eqn{kinst}, %
for which an $Nf\times Nf$ matrix must be diagonalized.
Other approaches for locating the instanton orbit are also naturally suggested
such as using the Hamilton-Jacobi formulation with end points constrained to bounce
\cite{GoldenRPI}
or modifications of the nudged-elastic-band method 
\cite{Einarsdottir2012path}.

Following \Ref{Althorpe2011ImF},
it can be shown that the semiclassical result \eqn{kGreens}
is equivalent to the RPI rate in the $N\rightarrow\infty$ limit \cite{rpinst}
and hence to the
standard instanton rate theories \cite{Coleman1977ImF,*Callan1977ImF,Affleck1981ImF,Benderskii}.
These %
are based on the \ImF\ premise, 
$	k Z_\text{r} \approx \frac{2}{\beta\hbar} \Im Z(\beta)$
\cite{Langer1969ImF,Weiss}
and the partition function can be evaluated in ring-polymer form as
\begin{align}
	\label{Fbeta}
	Z(\beta) &\equiv \eu{-\beta F} = \Lambda^{-Nf} \idotsint \eu{-\beta_N U_N(\mathbf{x})} \, \rmd\mathbf{x}.
\end{align}
Here, the integration is over $N$ ring-polymer beads $\mathbf{x}=\{\mat{x}_1,\dots,\mat{x}_N\}$;
$\beta_N=\beta/N$,
$\Lambda=\sqrt{2\pi\beta_N\hbar^2/m}$
and the ring-polymer potential is
\begin{align}
	U_N(\mathbf{x}) = \sum_{i=1}^N \frac{m}{2\beta_N^2\hbar^2}|\mat{x}_i-\mat{x}_{i+1}|^2 + V(\mat{x}_i),
\end{align}
where the indices are cyclic such that $\mat{x}_0\equiv\mat{x}_N$.
This is a discretization of the path-integral approach to quantum statistics \cite{Feynman},
and in the $N\rightarrow\infty$ limit, gives the partition function exactly.

The imaginary part of the partition function is, however,
not well defined
and
it can only be obtained using analytic continuation.
In practice,
one takes a steepest-descent integral about the saddle point of $U_N(\mathbf{x})$ \cite{rpinst,Althorpe2011ImF},
but reverses the sign of the negative eigenvalue
and multiplies the integral by a half \cite{Uses_of_Instantons,Kleinert}.
There is also a zero-eigenvalue mode that is integrated out analytically.
This procedure gives the RPI rate \cite{rpinst},
\begin{align}
	\label{kinst}
	k_\text{RPI} Z_\text{r}
	&= \frac{\Lambda^{-1}}{\beta_N\hbar} \sqrt{\scriptsize{\sum\nolimits_{i=1}^N|\mat{x}_i-\mat{x}_{i-1}|^2}} \sideset{}{'}\prod_k \left|\frac{1}{\beta_N\hbar\eta_k}\right| \eu{-\beta_N U_N}\!,
\end{align}
where $m\eta_k^2$ are the $Nf$ eigenvalues of the ring-polymer Hessian $\del^2 U_N$;
the prime indicates that the mode for which $\eta_k=0$ is not included in the product.

Although \eqn{kinst} is the form employed in RPI calculations,
equivalent expressions are found by taking the integrals in a different order \cite{Althorpe2011ImF}.
Steepest-descent integration of \eqn{Fbeta} over all beads but the two on the dividing surface gives
\begin{align}
	Z(\beta) &\simeq 2 \Lambda^{-2f} \iint \frac{1}{\sqrt{|\mathbf{J}^-| |\mathbf{J}^+|}} \, \eu{-\bar{S}(\mat{x}',\mat{x}'')/\hbar} \, \rmd\mat{x}' \rmd\mat{x}'',
\end{align}
where the factor of 2 appears because of the degeneracy of the ring-polymer space,
as the order of the beads along the orbit can be reversed.
The square Hessian matrices $\mathbf{J}^\pm$ %
are defined as in \Ref{GoldenRPI} from second-derivatives of $U_N(\mathbf{x})$
with respect to the beads on the $\pm\sigma>0$ side of the dividing surface.
A further coordinate transformation,
$\rmd\mat{x}'=\rmd q'\rmd\mat{Q}'=\dot{q}'\rmd\tau'\rmd\mat{Q}'$,
describes
the position along the trajectory using imaginary time.
The instanton orbit folds back on itself so
$\tau'$ has a range of $\half\beta\hbar$
and $\dot{q}'=\left|\der{q'}{\tau'}\right|$, %
which could be estimated using $|\mat{x}_{i+1} - \mat{x}_i|/\beta_N\hbar$ and the appropriate index $i$.
The equivalent holds for double primes.
Due to the cyclic permutational symmetry around the ring polymer \cite{rpinst},
the integral over one time variable is simple giving
\begin{align}
	Z(\beta)
	&\simeq 2 \Lambda^{-2f} \iiint \frac{\half\beta\hbar\dot{q}'\dot{q}''}{\sqrt{|\mathbf{J}^-| |\mathbf{J}^+|}} \, \eu{-\bar{S}/\hbar}
	\, \rmd\tau \rmd\mat{Q}' \rmd\mat{Q}'',
\end{align}
whereas the second over the remaining $\tau$
is completed, according to the usual \ImF\ procedure,
using analytic continuation of steepest-descent over an imaginary mode and multiplying by a factor of half:
\begin{align*}
	\Im Z(\beta) &\simeq \frac{\sqrt{2\pi\hbar}}{\Lambda^{2f}} \! \iint \! \frac{\half \beta\hbar\dot{q}'\dot{q}''}{\sqrt{|\mathbf{J}^-| |\mathbf{J}^+|}}  \left|\der[2]{\bar{S}}{\tau}\right|^{-\half} \! \eu{-\bar{S}/\hbar} \, \rmd\mat{Q}' \rmd\mat{Q}''.
\end{align*}
The remaining integrals
over the perpendicular directions
are performed using steepest-descent
to give 
\begin{align}
	\label{kImF}
	k_\text{RPI} Z_\text{r}
	&= (2\pi\hbar)^{-\half} \left(\frac{m}{\beta_N\hbar}\right)^{f} \frac{\dot{q}^2 |\Sigma|^{-1/2}}{\sqrt{|\mathbf{J}^-| |\mathbf{J}^+|}}
		\, \eu{-\bar{S}/\hbar},
\end{align}
where at the stationary point $\dot{q}'=\dot{q}''=\dot{q}$.
In the $N\rightarrow\infty$ limit,
this formulation %
is equivalent to all \ImF\ instanton rates \cite{Coleman1977ImF,*Callan1977ImF,Affleck1981ImF,Benderskii,rpinst,
Andersson2009Hmethane,*Andersson2011HCO,*Jonsson2011surface,
	DMuH,
	Goumans2010Hbenzene,*Goumans2011Hmethanol,*Meisner2011isotope,*Rommel2011locating,*Rommel2011grids,*Rommel2012enzyme,*Kaestner2013carbenes,*Kaestner2014review}
including \eqn{kinst}.

It is now a simple matter to show that \eqn{kImF} is equivalent to
the first-principles rate derived above from the semiclassical Green's functions,
i.e. $k_\text{SC}=\lim_{N\rightarrow\infty}k_\text{RPI}$.
From \eqn{Dbar},
and using a number of relations stated in \Refs{GoldenGreens,GoldenRPI},
the necessary equations are 
\begin{align}
	(-1)^{f+1} \Sigma^\pm / \dot{q}^2 
	&= \left|-\pders{\bar{S}^\pm}{\mat{x}'}{\mat{x}''}\right|
	= \left(\frac{m}{\beta_N\hbar}\right)^f |\mathbf{J}^\pm|^{-1}.
\end{align}

In summary,
the instanton method
for computing the rate of
tunneling through a barrier on a Born-Oppenheimer potential-energy surface
has been rederived from a semiclassical limit of scattering theory \cite{Miller1983rate}.
The final form is exactly equivalent to the usual expression given by the \ImF\ premise,
although the derivation is more rigorous.
The semiclassical instanton appears from the reaction probability at a given energy
before temperature has been introduced.
This is in contrast with other path-integral rate theories
based on the Boltzmann operator \cite{Gillan,Voth+Chandler+Miller,Makarov1995QTST,Cao1996QTST,Mills1997QTST,Habershon2013RPMDreview,Hele2013QTST}.
Real-time dynamical information does not contribute,
as is appropriate for a complex dissipative system where nuclear coherence is washed out.
However, unlike TST or QI methods \cite{Gillan,Voth+Chandler+Miller,Makarov1995QTST,Cao1996QTST,Mills1997QTST,Hele2013QTST,Miller2003QI,Vanicek2005QI},
the instanton rate remains independent of the dividing surface
so long as the instanton orbit intersects it.
In light of this new derivation,
applications of instanton methods
can be better understood
and the development of new RPMD and QI approaches advanced.
Generalizations of the new derivation provide a new route to solving novel problems
such as 
nonadiabatic reaction rates \cite{GoldenGreens}.
The author would like to thank Stuart C. Althorpe and William H. Miller for helpful comments on the manuscript.
This work was supported by the Alexander von Humboldt Foundation
and a European Union COFUND/Durham Junior Research Fellowship.

\appendix

\section{Appendix}

In the main text
it was claimed that the instanton partition function $Z^\ddag$
is equivalent to that given by Miller \cite{Miller1975semiclassical},
which is
expressed in terms of the
stability parameters, $u_j(E)$, of the instanton orbit \cite{Whittaker,Pars,Gutzwiller1971orbits,Miller1975semiclassical}.
This can be shown indirectly using the results from the main text
that the semiclassical rate, $k_\text{SC}$, is equivalent to the \ImF\ form, $k_\text{RPI}$,
and the proof in \Ref{Althorpe2011ImF} that $k_\text{RPI}$ is equivalent to Miller's rate \cite{Miller1975semiclassical}. %
However, it is also possible to provide a more direct proof
as outlined in this appendix.

The instanton was derived as the conjunction of two imaginary-time trajectories.
In order to make the connection with stability parameters, it will be necessary to make a transformation of the defining variables to
describe the instanton as a single periodic orbit.
The following derivation is similar to that followed in Section 4.4 of \Ref{Kleinert}.
The analysis applies equally to real-time and imaginary-time trajectories and the notation of a bar over the action is dropped here.

\begin{figure}
	\includegraphics{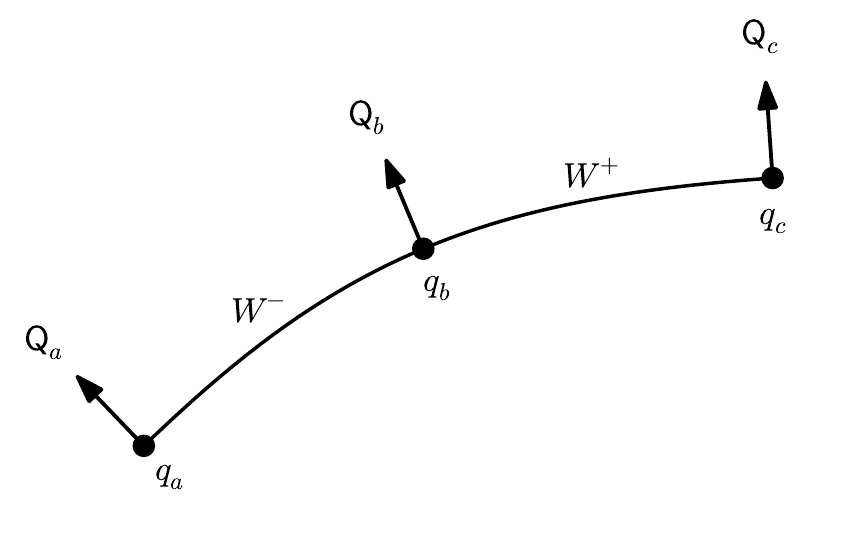}
	\caption{Schematic showing a classical trajectory
	traveling between points $(q_a,\mat{Q}_a)$ and $(q_c,\mat{Q}_c)$
	and which passes through $(q_b,\mat{Q}_b)$.
	The coordinates are parallel ($q$) and perpendicular ($\mat{Q}$) to the trajectory.
	The abbreviated action $W^\pm$ along each segment is marked.
	}
	\label{fig:connection}
\end{figure}

Consider first a classical trajectory with fixed energy, $E$,
traveling from $(q_a,\mat{Q}_a)$ to $(q_b,\mat{Q}_b)$ with abbreviated action $W^- \equiv W^-(\mat{Q}_a,\mat{Q}_b)$
and then continuing from $(q_b,\mat{Q}_b)$ to $(q_c,\mat{Q}_c)$ with abbreviated action $W^+ \equiv W^+(\mat{Q}_b,\mat{Q}_c)$.
The coordinate system is chosen such that the $q$ coordinate is parallel to the trajectory and $\mat{Q}$ perpendicular
and 
$q_a$, $q_b$ and $q_c$ are fixed.
In order that the two parts of the trajectory join correctly, 
$\mat{Q}_b=\mat{Q}_b(\mat{Q}_a,\mat{Q}_c)$
must be defined such that 
\begin{align}
	\label{join}
	\pder{W^-}{\mat{Q}_b} + \pder{W^+}{\mat{Q}_b} = \mat{0}.
\end{align}
Thus the two trajectories combine
to give one classical trajectory from $(q_a,\mat{Q}_a)$ to $(q_c,\mat{Q}_c)$
with abbreviated action %
\begin{align}
	\label{Wtilde}
	\tilde{W} \equiv \tilde{W}(\mat{Q}_a,\mat{Q}_c) = W^-(\mat{Q}_a,\mat{Q}_b) + W^+(\mat{Q}_b,\mat{Q}_c).
\end{align}
This situation is summarized in \fig{connection}.

\begin{widetext}

Partial differentiation of \eqn{Wtilde} using the chain rule gives
\begin{align}
	\pders{\tilde{W}}{\mat{Q}_a}{\mat{Q}_c} = \pders{W^-}{\mat{Q}_a}{\mat{Q}_b} \pder{\mat{Q}_b}{\mat{Q}_c} = \pders{W^+}{\mat{Q}_c}{\mat{Q}_b} \pder{\mat{Q}_b}{\mat{Q}_a}
\end{align}
and of \eqn{join} gives
\begin{subequations}
\begin{align}
	- \left.\pders{W^+}{\mat{Q}_b}{\mat{Q}_b}\right|_{\mat{Q}_c}
	&= \left.\pders{W^-}{\mat{Q}_b}{\mat{Q}_b}\right|_{\mat{Q}_c}
	= \pders{W^-}{\mat{Q}_b}{\mat{Q}_a} \pder{\mat{Q}_a}{\mat{Q}_b} + \left.\pders{W^-}{\mat{Q}_b}{\mat{Q}_b}\right|_{\mat{Q}_a},
\\
	- \left.\pders{W^-}{\mat{Q}_b}{\mat{Q}_b}\right|_{\mat{Q}_a}
	&= \left.\pders{W^+}{\mat{Q}_b}{\mat{Q}_b}\right|_{\mat{Q}_a}
	= \pders{W^+}{\mat{Q}_b}{\mat{Q}_c} \pder{\mat{Q}_c}{\mat{Q}_b} + \left.\pders{W^+}{\mat{Q}_b}{\mat{Q}_b}\right|_{\mat{Q}_c}.
\end{align}
\end{subequations}
In combination, they give
\begin{subequations}
\begin{align}
	\pders{\tilde{W}}{\mat{Q}_a}{\mat{Q}_a} &= \pders{W^-}{\mat{Q}_a}{\mat{Q}_a} - \pders{W^-}{\mat{Q}_a}{\mat{Q}_b} \left(\pders{W^-}{\mat{Q}_b}{\mat{Q}_b} + \pders{W^+}{\mat{Q}_b}{\mat{Q}_b}\right)^{-1} \pders{W^-}{\mat{Q}_b}{\mat{Q}_a}
\\
	\pders{\tilde{W}}{\mat{Q}_a}{\mat{Q}_c} &= - \pders{W^-}{\mat{Q}_a}{\mat{Q}_b} \left(\pders{W^-}{\mat{Q}_b}{\mat{Q}_b} + \pders{W^+}{\mat{Q}_b}{\mat{Q}_b}\right)^{-1} \pders{W^+}{\mat{Q}_b}{\mat{Q}_c}
\\
	\pders{\tilde{W}}{\mat{Q}_c}{\mat{Q}_a} &= - \pders{W^+}{\mat{Q}_c}{\mat{Q}_b} \left(\pders{W^-}{\mat{Q}_b}{\mat{Q}_b} + \pders{W^+}{\mat{Q}_b}{\mat{Q}_b}\right)^{-1} \pders{W^-}{\mat{Q}_b}{\mat{Q}_a}
\\
	\pders{\tilde{W}}{\mat{Q}_c}{\mat{Q}_c} &= \pders{W^+}{\mat{Q}_c}{\mat{Q}_c} - \pders{W^+}{\mat{Q}_c}{\mat{Q}_b} \left(\pders{W^-}{\mat{Q}_b}{\mat{Q}_b} + \pders{W^+}{\mat{Q}_b}{\mat{Q}_b}\right)^{-1} \pders{W^+}{\mat{Q}_b}{\mat{Q}_c},
\end{align}
\end{subequations}
and thus
\begin{multline}
	\pders{\tilde{W}}{\mat{Q}_a}{\mat{Q}_a} + \pders{\tilde{W}}{\mat{Q}_a}{\mat{Q}_c} + \pders{\tilde{W}}{\mat{Q}_c}{\mat{Q}_a} + \pders{\tilde{W}}{\mat{Q}_c}{\mat{Q}_c}
\\
	= \pders{W^-}{\mat{Q}_a}{\mat{Q}_a} + \pders{W^+}{\mat{Q}_c}{\mat{Q}_c} - \left(\pders{W^-}{\mat{Q}_a}{\mat{Q}_b} + \pders{W^+}{\mat{Q}_c}{\mat{Q}_b}\right)
			\left(\pders{W^-}{\mat{Q}_b}{\mat{Q}_b} + \pders{W^+}{\mat{Q}_b}{\mat{Q}_b}\right)^{-1}
			\left(\pders{W^-}{\mat{Q}_b}{\mat{Q}_a} + \pders{W^+}{\mat{Q}_b}{\mat{Q}_c}\right).
\end{multline}
\end{widetext}
These are the transformation equations for the general case of a trajectory split into two components.

The periodic instanton orbit is a special case of the trajectory considered above as its end points meet.
Using the notation from the main text,
it is defined with abbreviated action
\begin{align}
	W(\mat{Q}',\mat{Q}'') = W^-(\mat{Q}',\mat{Q}'') + W^+(\mat{Q}'',\mat{Q}'),
\end{align}
where
$q_a=q_b=q_c=0$, $\mat{Q}_a=\mat{Q}_c=\mat{Q}'$ and $\mat{Q}_b=\mat{Q}''$.

The instanton partition function, from \eqn{Zddag}, is therefore 
\begin{align}
	Z^\ddag &= 
	\sqrt{A^- A^+}
	\begin{vmatrix} \pders{W}{\mat{Q}'}{\mat{Q}'} & \pders{W}{\mat{Q}'}{\mat{Q}''} \\
				\pders{W}{\mat{Q}''}{\mat{Q}'} & \pders{W}{\mat{Q}''}{\mat{Q}''} \end{vmatrix}^{-\half}
\\
	&= \sqrt\frac{\Psi}{\Phi}
\end{align}
where
\begin{align}
	\Phi
	&= \left|\pders{W}{\mat{Q}'}{\mat{Q}'} - \pders{W}{\mat{Q}'}{\mat{Q}''} \left(\pders{W}{\mat{Q}''}{\mat{Q}''}\right) \pders{W}{\mat{Q}''}{\mat{Q}'} \right|
\\
	&= \left|\pders{\tilde{W}}{\mat{Q}_a}{\mat{Q}_a} + 2\pders{\tilde{W}}{\mat{Q}_a}{\mat{Q}_c} + \pders{\tilde{W}}{\mat{Q}_c}{\mat{Q}_c} \right|
\end{align}
and
\begin{align}
	\Psi &= A^- \left|\pders{W}{\mat{Q}''}{\mat{Q}''}\right|^{-1} A^+ 
\\
	&= \left| \pders{W^-}{\mat{Q}_a}{\mat{Q}_b} \left(\pders{W^-}{\mat{Q}_b}{\mat{Q}_b} + \pders{W^+}{\mat{Q}_b}{\mat{Q}_b}\right)^{-1} \pders{W^+}{\mat{Q}_b}{\mat{Q}_c} \right|
\\
	&= \left| - \pders{\tilde{W}}{\mat{Q}_a}{\mat{Q}_c} \right|
\end{align}

Section 4 of \Ref{Gutzwiller1971orbits} shows that the ratio of these determinants gives
\begin{align}
	Z^\ddag
	&= \prod_{j=1}^{f-1} \frac{1}{2\sinh[u_j(E)/2]}
\end{align}
where $u_j(E)$ are the non-zero stability parameters of the periodic orbit.
It is known that the stability parameters do not depend on the position $q_a=q_c$ around the orbit \cite{Gutzwiller1971orbits}
and thus it is proved that, as stated in the main text,
the semiclassical reaction probability is independent of the form of the dividing surface.
The same stability parameters also appear in Miller's instanton theory \cite{Miller1975semiclassical}
which is therefore equal to the semiclassical rate derived in the main text.

\end{document}